\documentclass[twocolumn,dvipsnames,twocolappendix]{aastex631}
\usepackage{amsmath, amsfonts, amssymb, graphicx, chngcntr, multirow, float}
\usepackage[]{hyperref}
\hypersetup{
	colorlinks=true,
	urlcolor=MidnightBlue,
	linkcolor=WildStrawberry,
	citecolor=ForestGreen,
}

\newcommand{\XMM}{{\it XMM-Newton}}
\newcommand{\Swift}{{\it Swift}}
\newcommand{\Kepler}{{\it Kepler}}
\newcommand{\qats}{\texttt{QATS}}

\received{September 28, 2021}
\revised{October 17, 2021}
\accepted{\today}
\submitjournal{ApJL}

\shorttitle{J0249 QPEs}
\shortauthors{Chakraborty \textit{et al.}}

\begin{document}

\title{Possible X-ray Quasi-Periodic Eruptions in a Tidal Disruption Event Candidate}

\author[0000-0002-0568-6000]{Joheen Chakraborty}
\affiliation{Department of Astronomy, Columbia University, 550 W 120th Street, New York, NY 10027, USA}
\affiliation{MIT Kavli Institute for Astrophysics and Space Research, Cambridge, MA 02139, USA}

\author[0000-0003-0172-0854]{Erin Kara}
\affiliation{MIT Kavli Institute for Astrophysics and Space Research, Cambridge, MA 02139, USA}

\author[0000-0003-4127-0739]{Megan Masterson}
\affiliation{MIT Kavli Institute for Astrophysics and Space Research, Cambridge, MA 02139, USA}

\author[0000-0002-1329-658X]{Margherita Giustini}
\affiliation{Centro de Astrobiología (CSIC-INTA), Camino Bajo del Castillo s/n, Villanueva de la Cañada, 28692 Madrid, Spain}

\author[0000-0003-0707-4531]{Giovanni Miniutti}
\affiliation{Centro de Astrobiología (CSIC-INTA), Camino Bajo del Castillo s/n, Villanueva de la Cañada, 28692 Madrid, Spain}

\author[0000-0002-4912-2477]{Richard Saxton}
\affiliation{XMM–Newton SOC, ESAC, Apartado 78, E-28691 Villanueva de la Cañada Madrid, Spain}

\correspondingauthor{Erin Kara}
\email{ekara@mit.edu}

\begin{abstract}

 X-ray Quasi-Periodic Eruptions (QPEs) are a recently discovered phenomenon associated with supermassive black holes at the centers of galaxies. They are high amplitude soft X-ray flares that recur on timescales of hours, but what causes these flares remains uncertain. In the two years since their original discovery, four known QPE-hosting galaxies have been found, with varying properties and levels of activity. We have conducted a blind algorithm-assisted search of the \XMM\ Source Catalog and found a fifth QPE candidate, XMMSL1 J024916.6-041244. This is a star-forming galaxy hosting a relatively low-mass nuclear black hole, and has previously been identified as a Tidal Disruption Event candidate. An \XMM\ pointed observation of the source in 2006 exhibited nearly two QPE-like flares in soft X-rays, and, unlike in other QPE sources, there are hints of corresponding dips in the UV light curves. Afterwards, a series of \Swift\ observations observed the rapid dimming of the source; thereafter, in August 2021, we triggered a second \XMM\ observation, which revealed that the source is detected, but the QPEs are no longer present. Here we report on (I) the strategy we used to systematically search through \XMM\ archival data; (II) the properties of J0249 and its QPE flares; and (III) the relative behaviors and properties of the QPE sample to date, now 5 members large.
\end{abstract}


\section{Introduction} \label{sec:intro}
Quasi-periodic eruptions (QPEs) are a recently discovered class of rapid soft X-ray flares (typically concentrated in the $0.5-2$ keV band) originating from the nuclei of both active and inactive galaxies hosting low-mass central supermassive black holes (SMBHs). Four systems exhibiting QPEs have been found to date: the first two, GSN 069 and RX J1301.9+2747 \citep{Miniutti19,Giustini20}, are consistent with active galactic nuclei (AGN) on the basis of their emission line profiles, though it is worth noting that both galaxies show no broad components to the $H\alpha$ or $H\beta$ lines, and that the observed narrow emission lines may be the signature of past nuclear activity. The latter two sources, eRO-QPE1 and eRO-QPE2 \citep{Arcodia21}, are quiescent galaxies, and may be more representative of the overall QPE population, given the blind-search nature of the \textit{eROSITA} survey used to find them.

 The QPE flares of all four previously known systems share a number of unique observational properties which distinguish them from typical AGN variability, indicating that they likely share a common physical origin. The bursts have durations on the timescale of $\sim$a few kiloseconds, with an average peak-to-peak separation time of tens of kiloseconds. Within each source, the remarkably symmetric flares have higher amplitudes, shorter durations, and earlier peaks in higher energy bands \citep{Miniutti19,Giustini20}. Correspondingly, phase-resolved spectroscopy of QPE systems shows that their very soft spectra are harder during high-flux states.

On the basis of these observational features, we conducted a blind search of archived data collected by the \XMM\ telescope \citep{Jansen01} to search for evidence of more QPE-exhibiting systems. Rather than limiting to a narrow class of host galaxy physical properties, we scanned the entire archive using a computer algorithm-aided approach (Sec.~\ref{sec:methods}), and found one such system: XMMSL1 J0249-041244, hereafter J0249.

J0249 was first detected in 2004 by the \XMM\ Slew Survey and followed up with an 11.7~ks pointed \XMM\ observation in 2006 \citep{Esquej07}. It was first identified as a TDE candidate due to a rise in flux by a factor of 88 compared to a ROSAT upper limit \citep{Strotjohann16}. As is commonly seen in X-ray TDEs, its spectrum is very soft, and is well described by a blackbody, with no additional hard X-ray emission. The galaxy was initially classified as a Seyfert 1.9 TDE host candidate \citep{Esquej07}, but more recently higher resolution spectra led to its reclassification as a star-forming galaxy potentially hosting an AGN \citep{Wevers19}. The mass of the central SMBH is highly uncertain. \cite{Wevers19} reported a mass of $\sim 8.5 \times 10^4 M_\odot$ as inferred using bulge velocity dispersions from absorption lines and the $M-\sigma$ relation (which has large scatter in this low-mass regime), while \cite{Strotjohann16} reported an estimated mass of $\sim 5 \times 10^5 M_\odot$ based on the K-band luminosity and the empirical $M_{BH}-M_*$ relation; the order-of-magnitude discrepancy in reported sizes suggests large systematic uncertainties on the mass estimate. Since its discovery in 2004 until 2017, J0249 was observed about 15 times with both XMM-Newton and Swift, revealing a gradual long-term dimming by over an order of magnitude and generally following a $t_{yrs}^{-5/3}$ behavior (Fig.~\ref{fig:lc}), as expected for the fallback rate of a TDE \citep{Rees88}. During the 2006 \XMM\ exposure, J0249 exhibited 1.5 symmetric QPE-like flares, increasing in X-ray luminosity from a quiescent level of $L_{0.5-2}=1.6\times 10^{41}$ erg s$^{-1}$ to a flare level of $L_{0.5-2}=3.4\times 10^{41}$ erg s$^{-1}$. The maximum luminosity increase is 11.2x in the 1-1.3 keV band. In this letter, we report on the properties of this QPE-like flare.

In Section~\ref{sec:methods}, we discuss the approach used to search through \XMM\ archival data to identify and analyze promising QPE candidates. In Section~\ref{sec:results} we discuss the results from our analysis of data from both \XMM\ observations of J0249. In Section~\ref{sec:discussion} we make some remarks on the QPE population to date.

\section{Observations and Methods} \label{sec:methods}
\begin{table*}
\caption{\XMM\ EPIC-PN logs for the 2006 (Full Frame, Medium optical filter) and 2021 (Full Frame, Thin optical filter) observations of J0249.}
\centering
\begin{tabular}{c c c c c c c}
\hline
\hline
 OBSID & Start date & Exposure & (src+bkg)$_{0.3-2}$ & (bkg)$_{0.3-2}$ & (src+bkg)$_{2-10}$ & (bkg)$_{2-10}$ \\
   & (yyyy-mm-dd hh:mm:ss) & (s) & (counts s$^{-1}$) & (counts s$^{-1}$) & (counts s$^{-1}$) & (counts s$^{-1}$) \\
\hline
0411980401 & 2006-07-14 11:02:44 & 11739 & $0.325 \pm 0.01$ & $0.006 \pm 0.0008$ & $0.012 \pm 0.002$ & $0.008 \pm 0.001$ \\
0891800601 & 2021-08-06 17:02:02 & 33800 & $0.015 \pm 0.001$ & $0.004 \pm 0.0008$ & $0.0083 \pm 0.0006$ & $0.005\pm 0.001$ \\
\hline
\end{tabular}
\label{tab:xmmobs}
\end{table*}

\subsection{Quasi-periodic Automated Transit Search (\qats)}
Our algorithm of choice for searching through the \XMM\ archive was the Quasi-periodic Automated Transit Search (\qats). The algorithm, originally developed in \cite{Carter13}, was designed to find exoplanet transit timing variations (TTVs) in \Kepler\ optical data. \qats\ is a maximum-likelihood algorithm which models a candidate transit at each feasible cadence in a light curve, then compares the $\chi^2$ fit to a polynomial continuum representing the baseline, identifying quasi-periodic signals where the transit fit outperforms the continuum. Apart from TTVs, \qats\ has also been used to find ``inverted transit" systems, e.g. self-lensing binary stars \citep{Kruse14} showing quasi-periodic symmetric brightenings rather than dimmings, similar to the behavior of QPEs. \qats\ thus provides an attractive option for en-masse triaging of \XMM\ archival light curves.

We made use of data from the 4XMM \XMM\ serendipitous source catalog compiled by the 10 institutes of the \XMM\ Survey Science Center (SSC) selected by ESA \citep{Webb20}. Preprocessed 0.2-12 keV light curves from the \XMM\ SSC pipeline were retrieved directly from the web interface\footnote{\href{http://xmm-catalog.irap.omp.eu}{http://xmm-catalog.irap.omp.eu}} developed by \cite{Zolotukhin17}. In total, this consisted of 302,773 broadband light curves taken from 11,647 observations made during 2000-2019. We then ran \qats\ on all of these light curves and sorted them by the \qats\ merit function $S$, a quantity derived from the Gaussian log-likelihood. The \qats\ merit function can be interpreted by its relation to the signal-to-noise ratio, $S \equiv \sigma \times (S/N)_{\mathrm{total}}$. High-performing signals were then vetted by-eye and then reduced and analyzed manually.

\subsection{Data Reduction}
``Promising" candidates from our \qats\ search pipeline (i.e. showing one or multiple high-amplitude variability events separated by stable quiescent periods in their broadband light curves) were subsequently reduced and analyzed using the \XMM\ Science Analysis System (SAS) v18.0.0, following the standard SAS threads recommended by the \XMM\ Science Operations Center. Spectral fitting of EPIC-pn data was performed using HEASoft v6.28 with Xspec v12.11.1. In almost all cases, false-positives were vetted out during this stage (roughly 40 total), for reasons including (I) lack of spectral hardening during flare states; (II) energy-resolved light curves showing flares not in agreement with the QPE energy dependence, i.e. higher amplitude and smaller duration in higher energy bands, or flares not isolated to a narrow energy range; or (III) excessive soft proton background flaring indicating that observed flux variability is unlikely to be confined to a central source. Common false-positive sources include X-ray binaries and stochastic variability from AGNs/quasars. Importantly, publicly archived \XMM\ observations of GSN 069 and RX J1301.9+274 exhibiting QPE flares were recovered by this process. Ultimately, J0249 was the only novel candidate which passed all of our false-positive tests.

The \Swift\ data used to produce its long-term light curve (Fig.~\ref{fig:lc}) were analyzed with the online processing tool maintained by the University of Leicester\footnote{\href{https://www.swift.ac.uk/user_objects}{https://www.swift.ac.uk/user\_objects}}.

\subsection{Observations of J0249}
After the initial detection from the \XMM\ Slew Survey, J0249 was observed again in 2006 with an \XMM\ pointed observation lasting 9.9 ks in the EPIC-PN detector and 11.7 ks in the EPIC-MOS detectors (OBSID: 0411980401). This is the observation initially flagged by the QATS algorithm.

\begin{figure*}
\centering
\includegraphics[width=\textwidth]{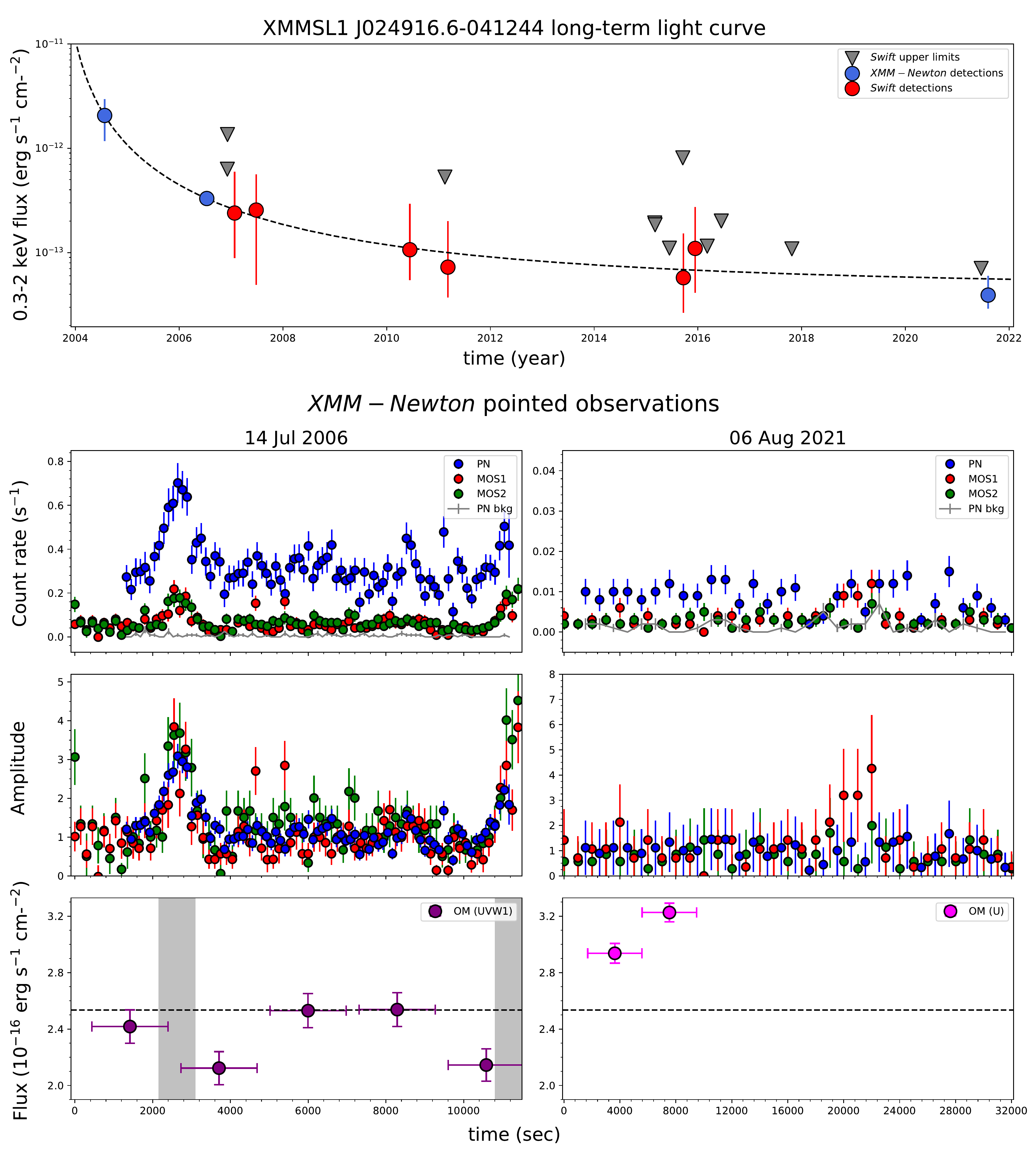}
\caption{\textbf{Top:} 0.3-2 keV flux evolution of J0249 since the initial \XMM\ slew detection in 2004, with \XMM\ and \Swift\ detections fit by a $t_{\mathrm{yrs}}^{-5/3}$ power law decay model as expected of fallback from a TDE \citep{Rees88}. \textbf{Bottom:} Background-corrected 0.3-2 keV light curves from the 2006 and 2021 \XMM\ pointed observations. Time bins are 120/150 seconds for 2006 PN/MOS data and 1000 seconds for 2021 (the coarse binning of the 2021 light curve is chosen to reduce error bar size). Note the different y-axis ranges for the non-normalized X-ray light curves of 2006 and 2021, as well as the different x-axis binning from the longer exposure length in 2021. Error bars are plotted at 90\% confidence. Amplitudes are normalized to median quiescent count rate. \XMM\ Optical Monitor data obtained using the UVW1 (2006) and U (2021) filters are also shown, with vertical bars denoting 90\% confidence intervals and horizontal bars denoting time bins. The dashed line indicates the 2006 quiescent flux level for comparison, while shaded regions indicate flare periods}.
\label{fig:lc}
\end{figure*}

The source was observed 15 times from 2006-2017 with the \textit{Swift} XRT, revealing a gradual long-term dimming by over an order of magnitude and generally following a $t_{yrs}^{-5/3}$ behavior (Fig.~\ref{fig:lc}), as expected for the fallback rate of a TDE \citep{Rees88}. We also attempted to fit the dimming by a $t_{yrs}^{-9/4}$ model as expected of the fallback rate from a partial tidal disruption event \citep{Miles20}, but this fit was considerably poorer.

After our initial discovery of the source as a QPE candidate, we requested a 5 ks \textit{Swift} Target-of-Opportunity observation, which was carried out in June 2021 and resulted in an upper limit consistent with the observed dimming . We then requested a longer 33.8 ks \XMM\ Director's Discretionary Time (DDT) observation that was performed on August 6, 2021 (OBSID: 0891800601), which revealed that the flares were no longer present but the source was still detectable. As only 1.5 flares in total were detected from the source, the classification of J0249 as a true QPE source is less clear than previous ones. However, given that the characteristics of the flares and quiescence align closely with known QPEs, and that these flares are distinct from those seen in other channels for AGN variability, we refer to the source as a QPE candidate.

\section{Results} \label{sec:results}

\subsection{Light curve analysis}

During the 2006 observation, 1.5 symmetric flares separated by 9 ks and confined almost entirely to the 0.8-2 keV band were detected in the EPIC-PN, MOS1, and MOS2 \citep{Strueder01, Turner01} light curves (Fig.~\ref{fig:lc}, Fig.~\ref{fig:eresolved}). The 0.5-2 keV X-ray luminosity increased from a quiescent level of $L_{0.5-2}=1.6\times 10^{41}$ erg s$^{-1}$ to a flare level of $L_{0.5-2}=3.4\times 10^{41}$ erg s$^{-1}$.  Assuming a black hole mass of $8.5 \times 10^4 M_\odot$, the quiescent Eddington ratio $R_{\mathrm{Edd}}$ is $\approx 0.13$. Correspondingly, \cite{Wevers19} found an average peak $R_{\mathrm{Edd}}$ of 0.27 among their sample of soft X-ray detected TDE candidates including J0249.

Similar to the four other QPE sources, the light curve shows a relatively stable quiescent flux apart from these rapid variability events. By the August 2021 \XMM\ pointed observation, the dimming of the source had resulted in a flux decrease of over an order of magnitude. While the 2021 \XMM\ observation was designed to be long enough to catch 2--3 QPE flares, there is no longer significant X-ray variability, meaning that within the 15 years after the original QPE detections, the phenomenon has ceased (Fig.~\ref{fig:lc}).

In order to quantify the QPE duration and recurrence time, we model the light curves of the 2006 \XMM\ observation using a constant baseline equal to the mean quiescent count rate, and represent the symmetric QPE flares using Gaussians. The QPE amplitudes, peaking times and durations correspond to the Gaussian amplitude, centroid and FWHM, respectively. For the first flare detected in the 2006 observation, as we probe higher energy bands from 0.3-1.3 keV, the amplitudes increase greatly, durations generally decrease moderately, and peaking times generally decrease slightly (Fig.~\ref{fig:eresolved}). We refrain from quoting flare amplitudes in the 1.3-2 keV band due to poor S/N. As only part of the second flare was seen by all three cameras, it is unclear what its amplitude, peaking time, or durations were. Above 2 keV, the background dominates, and no flaring behavior is seen. The energy dependence of the flare properties align with trends seen in the other QPE sources.

The UVW1 filter of the \XMM\ optical monitor (OM) instrument \citep{Mason01} also shows flux variability during the 2006 pointed observation, though it is not strictly coincident with the soft X-ray flares, perhaps due to the long UVW1 exposure time compared to the X-ray time binning. Lower flux states are seen shortly preceding and following the first X-ray flare, with a third lower flux exposure aligned with the beginning of the second flare. During the 2021 observation, the first two OM exposures used the U filter, resulting in two detections of the source; however, subsequent OM exposures used the UVW2 filter and did not detect J0249.

\begin{figure*}
\centering
\includegraphics[width=\textwidth]{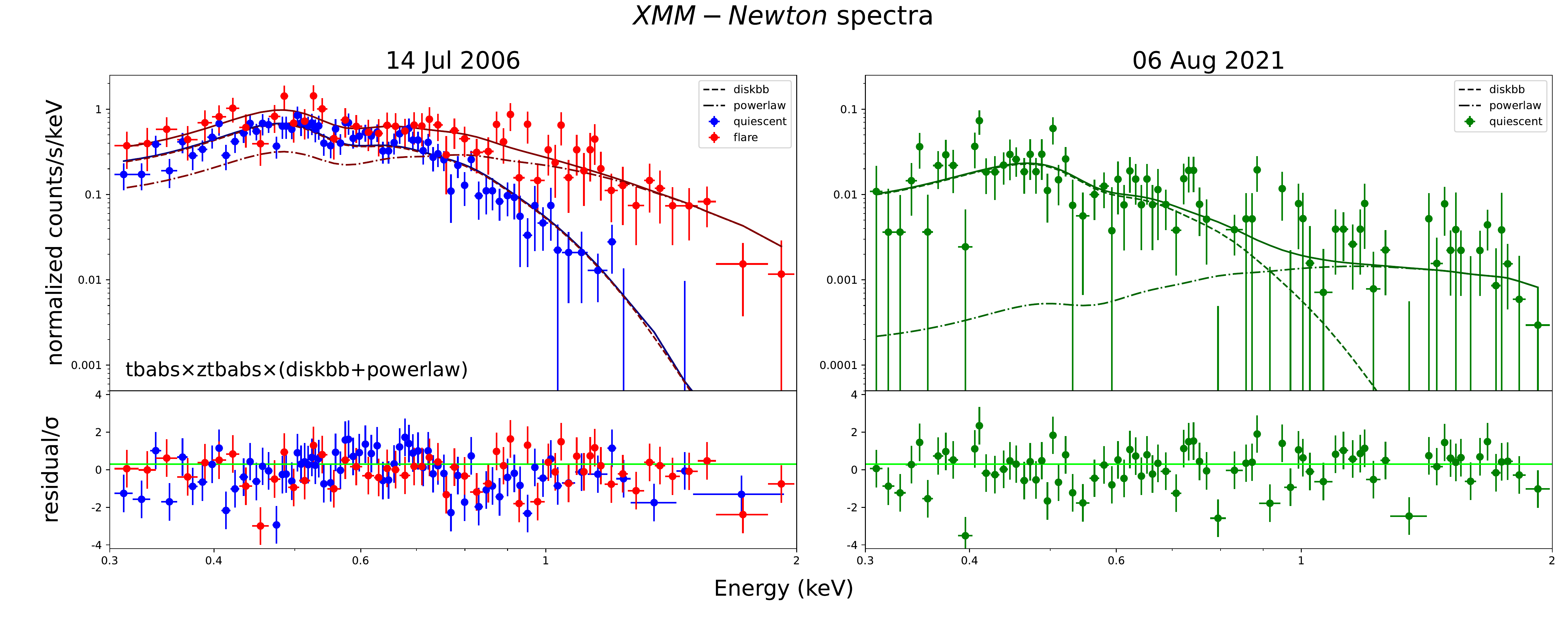}
\caption{Flux-resolved EPIC-PN spectra of J0249 for both \XMM\ pointed observations, along with \texttt{tbabs$\times$ztbabs$\times$(diskbb+powerlaw)} model fits (Table~\ref{tab:specfit}). Error bars are plotted at 90\% confidence. Note the different y-axes for the two observations resulting from the large flux difference of J0249 over the elapsed period. The 2006 observation is separately fit during the quiescent and flaring phases with $N_H(z)$ and blackbody temperature \& normalization tied. The observed spectral variability in 2006 closely matches what is seen in other QPE candidates, i.e. fast transitions from a disk-dominated quiescent phase to a state with harder emission from additional hot component.}
\label{fig:spec}
\end{figure*}

\subsection{Spectral analysis}

We perform flux-resolved spectral analysis on the 2006 0.3-2 keV EPIC-PN data (because of its higher $S/N$ and hence time resolution compared to the MOS detectors). We divide the data into flare and quiescent states using a threshold of 0.4 cts sec$^{-1}$ to separate low and high flux. For the 2006 observation we group the background-subtracted spectra with a minimum of 1 count per energy bin, and use the Cash statistic \citep{Cash79} to fit various models to the data. All spectral analysis results are reported in Table~\ref{tab:specfit}. The most notable result from spectral fitting is that, similar to other QPEs, we see a hardening of the X-ray spectrum during flaring states.

We also analyze the spectra of the 2021 0.3-2 keV EPIC-PN data, but do not resolve into separate high- and low-flux phases due to the lack of variability from the QPE ``turn-off" between 2006 and 2021. The decrease in flux of J0249 to the level of $\sim 4\times 10^{-14}$ erg cm s$^{-1}$ makes constraining spectral parameters difficult due to low SNR, but for completeness we nonetheless report results of spectral fitting here. As with the 2006 data, we group with 1 count per energy bin and fit using the Cash statistic.

In all fits, we model galactic line-of-sight absorption using the \texttt{tbabs} Xspec model \citep{Wilms00}. \cite{Strotjohann16} found that statistically acceptable fits of J0249 favor high intrinsic $N_H$, which we model using \texttt{ztbabs}. For our different fits, we test various combinations of blackbodies (\texttt{diskbb} and \texttt{bbody}) and powerlaws.

Model 1, \texttt{tbabs$\times$ztbabs$\times$diskbb} (Tab.~\ref{tab:specfit}), is the simplest model for which we obtain reasonable spectral fits, with a reduced $\chi^2 \leq 1.31$ for 325 and 145 degrees of freedom corresponding to 2006 and 2021 spectra, respectively. We allow for the blackbody temperature and normalization to vary between quiescent and flare periods in the 2006 observation, but require that they have the same intrinsic $N_H$. Interestingly, the change in disk temperature during the high-flux state does not follow $L \propto T^4$ emission as expected from the Stefan-Boltzmann Law; this pattern is also seen in other QPE sources.  For example, the Model 1 fit of J0249 has $(kT_{flare}/kT_{quiescent})^4 = 6.2$, but only $L_{flare}/L_{quiescent} = 2.08$. This motivates the use of a more complex model, where changes are not simply due to the disk blackbody.

We test a model where the disk blackbody temperature and normalization remains constant from quiescent to flare periods, but where the QPE flares are due to the presence of an additional harder component. In Model 2, we model this harder component as a powerlaw: \texttt{tbabs$\times$ztbabs$\times$(diskbb+powerlaw)}  (Fig. ~\ref{fig:spec}). During the quiescent phase, the normalization of the powerlaw component is consistent with zero, favoring a pure disk spectrum, whereas during the flare state addition of this component results in considerably better fit statistic with a reduced $\chi^2 < 1.13$. The photon index of the additional powerlaw is extremely soft, and does not behave like a regular AGN hot corona ($\Gamma \sim 1.8$).

We then test whether the harder component is better described by a second, hotter blackbody, rather than a soft powerlaw: \texttt{tbabs}$\times$\texttt{ztbabs}$\times$(\texttt{diskbb}+\texttt{bbody}) (Model 3). Again, as in Model 2, we keep the lower temperature disk blackbody tied between quiescent and flaring periods. Comparing to Model 2, Model 3 provides a better statistical fit for the 2006 observation, but a slightly worse fit for the 2021 observation. Additionally, the secondary blackbody normalization during quiescence is also consistent with zero, confirming the finding from Model 2 that a pure disk spectrum is the favored model during the quiescent phase.

Model 1 is then repeated using \texttt{zxipcf} in place of \texttt{ztbabs} to explore the effect of assuming ionized instead of neutral gas. Part of the motivation behind this model is the residuals around 0.7-0.8 keV in the previous models, which have been speculated as being due to OVII or OVIII absorption features from an outflow \citep{Brandt97}. Similar absorption-like features have been seen in the soft X-ray spectra of TDEs (e.g. ASASSN-14li; \citealt{kara18}). For this fit, we leave the internal column density $N_H(z)$ free to vary and fix $\log(\xi)=2.95$, a choice motivated by \cite{Strotjohann16}. The model performs worse than the previous three (reduced $\chi^2 \leq 1.52$), leaving the ionization features of the internal gas in J0249 uncertain. Attempting a fit of multiple emitters alongside the disk using ionized gas (e.g. \texttt{tbabs$\times$zxipcf$\times$(diskbb+bbody)}) does not result in significant improvement in the fit statistic.

We conclude from the fitting results that the high- and low-flux spectra are best described by the same non-variable disk component, while the QPE flares are due to an additional hotter component described by either a second blackbody emitter or a powerlaw behavior. This additional component does not behave like a regular AGN hot corona with $\Gamma \sim 1.8$, but is instead much softer, as is typical of the AGN soft excess. Finally, the harder component persists in 2021, but now the photon index is harder ($\Gamma=1.8\pm1.6$), which is more aligned with expectations of AGN hot coronae. The late time hard tail (perhaps the late time appearance of a hot corona) has been seen in a few X-ray TDEs (e.g. \citealt{kara18}; \citealt{saxton20} for a review) and Changing-Look AGN (e.g. \citealt{ricci20}).

\section{Discussion} \label{sec:discussion}
\begin{figure*}
\centering
\includegraphics[width=\textwidth]{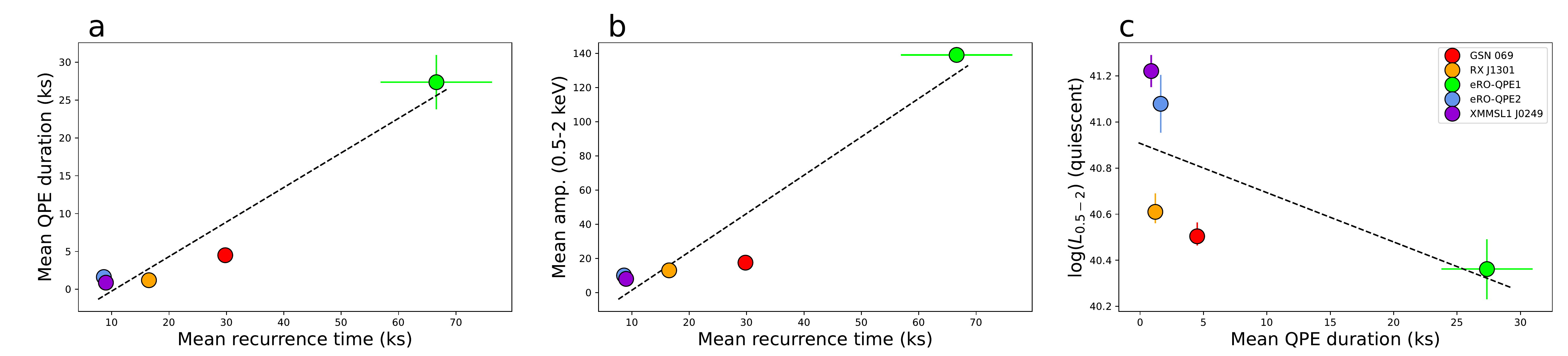}
\caption{Comparison of various physical and observational properties of the five known QPE sources to date.}
\label{fig:qpepop}
\end{figure*}
 
We present the discovery of the 5th QPE candidate thus far, in the low-mass star-forming/composite galaxy, XMMSL1 J024916.6-041244, which has previously been identified as a likely TDE candidate \citep{Esquej07, auchettl18, Wevers19}. Our major findings are:
\begin{itemize}
    \item 1.5 QPE-like soft X-ray flares were discovered in the brightest XMM-Newton pointed observation.
    \item For the first time, we find tentative evidence for corresponding non-X-ray variability in the form of UV flux dips coincident with the X-ray flares.
    \item As with all other QPE candidates, the spectra are consistent with a picture in which the disk blackbody remains constant, but the QPE-like flares are due to the emergence of an additional `soft excess' component.
    \item The QPE-like flares disappear at late times, after the source has dimmed considerably. 
    \item The late time (post-QPE) spectra show a harder spectrum, consistent with the emergence of a hot corona, which is typical of normal AGN.
\end{itemize}

\subsection{The QPE population to date}
J0249 represents the fifth member of a growing QPE population. Here we consider the relationships between various physical and observed flaring properties (quiescent $0.5-2$ keV luminosity $L_X$, flare amplitude, flare duration, and flare recurrence time $t_{\mathrm{rec}}$) of QPE sources.

Fig.~\ref{fig:qpepop} shows scaling relationships of mean $t_{\mathrm{rec}}$ with mean duration and mean amplitude; data are taken directly from the original discovery papers \citep{Miniutti19,Giustini20,Arcodia21}, with error bars provided where possible. We note that these quantities, particularly amplitude, can be highly variable even within individual QPE sources---in the most extreme case of eRO-QPE1, the amplitudes of different flares can differ by a factor of up to 10x. Within sources showing high variability of QPE amplitude (RX J1301.9+2747, eRO-QPE1), a shorter $t_{rec}$ is associated with a larger amplitude, the inverse of the trend seen between separate QPE sources. QPE sources also show alternating ``long-short" recurrence times associated with respective smaller and larger amplitude flares.

Recent work has explored the possibility of QPEs being generated by accretion from orbiting bodies such as extreme-mass ratio inspirals/EMRIs \citep{Arcodia21, Metzger21} or the partial tidal disruptions of a star \citep{King20}. These scenarios provide compelling arguments for the quasi-periodic nature of QPEs by allowing for some residual orbital eccentricity to modulate the reccurence time ($t_{rec}$) and duration of the flares. They would also account for the inverse relationship seen between the X-ray luminosity $L_X$ and characteristic QPE timescales in Fig.~\ref{fig:qpepop}{\color{WildStrawberry}c}, as a larger accretion rate (which scales closely with $L_X$ in this low-mass SMBH regime) leads to a thicker disk, and thus shorter $t_{rec}$ and duration.

J0249 falls near the shorter end of QPE characteristic timescales, closely matching the durations and recurrence times seen in eRO-QPE2 in spite of its host galaxy sharing the most physical similarities with GSN 069, a comparatively intermediate QPE source. eRO-QPE1 is a conspicuous outlier compared with the four other QPE hosts in terms of characteristic timescales, quiescent luminosity, flare amplitude, and blackbody temperature. Discovery of further QPE hosts in the intermediate regime would provide an important information on the scaling relationships illustrated in Fig.~\ref{fig:qpepop}.

\subsection{The UV variability}

J0249 is unique among QPE sources in that it shows evidence for dips in the UV around the same time as the X-ray flares. Due to the low number of observed flares in J0249, we cannot claim a definitive correlation with the UV, but the light curves are suggestive. \citet{Arcodia21} showed that in the two eROSITA QPEs, there was no corresponding variability in the optical/UV. Similarly, no significant UV activity was seen in GSN~069 \citep{Miniutti19}, though a re-analysis of RX J1301.9+2747 from the May 2019 observation did show that one of the three X-ray flares is accompanied by a slight dip in UV (but much smaller amplitude than in J0249). 

\citet{Arcodia21} suggest that the lack of UV/optical variability may be due to particularly small accretion disks in these two previously quiescent galaxies. Coincident UV activity in J0249 may instead suggest that the QPE phenomenon also occurs on larger scales of $\sim$thousand gravitational radii from the black hole. Or perhaps, since the host allows for the presence of an AGN, J0249 may have had a large pre-existing accretion disk that couples to the QPE phenomena at small scales. Future observations of QPEs in both quiescent galaxies and known AGN will elucidate these findings.

\section{Conclusion} \label{sec:conclusion}
During an 11.7 ks observation performed by \XMM\ in July 2006, 1.5 rapid symmetric QPE-like flares separated by 9 ks were observed from the star-forming/composite galaxy XMMSL1 J024916.6-041244. We call these flares QPE-like because they share with previously known QPEs the following set of distinctive properties: (I) they are confined to soft energy bands ($<2$ keV); (II) they are higher-amplitude and shorter-lived in higher energy bands; (III) the high-flux states are associated with rapid transitions to harder spectra; and (IV) the quiescent intervals between flares are remarkably stable, thus indicating the flares are unrelated to other forms of traditional AGN variability.

After the initial detection, the source dimmed considerably, by over an order of magnitude in the period from 2006 to 2021. Our follow-up 33.8 ks \XMM\ DDT observation in August 2021 found the source in a low-flux, non-variable state, indicating that QPE-like activity had ceased in J0249 altogether. Recent suggestions for the physical mechanism behind QPE behavior invoke systems of multiple objects, e.g. black hole binaries or orbiting companions around a central black hole, with the flares being produced by periodic gravitational lensing events \citep{Ingram21} or accretion from an orbiting companion with high-flux states corresponding to recurring phases of amplified mass transfer \citep{King20, Metzger21}. If these models are the true source of QPEs, it is possible that shortly after 2006 we witnessed the infall and capture of the orbiting companion by the central SMBH of J0249; this would explain both the rapid dimming of the source, as well as the lack of any QPE flares now.

Along with GSN 069, J0249 marks the second potential QPE source discovered by the \XMM\ slew survey. Compared with a total of 10 soft-X-ray nuclear transients discovered by the same survey \citep{saxton20} with sufficiently long observations to detect QPEs, we might expect a rough lower limit of 20\% of soft-X-ray nuclear transient sources showing QPEs. Of course, given the small-number statistics, this estimate carries large uncertainty. \\

We thank \XMM\ Project Scientist Norbert Schartel for approving our DDT request. JC and EK are supported by NASA grant 80NSSC20K1084. This research is based on observations obtained with \XMM, an ESA science mission with instruments and contributions directly funded by ESA Member States and NASA.

\bibliography{refs}{}
\bibliographystyle{aasjournal}

\appendix
\counterwithin{figure}{section}
\counterwithin{table}{section}
\section{Supplementary figures and tables}
\begin{table*}
\caption{EPIC-PN spectral analysis results. Errors are quoted at 90\%   confidence. For Models 2 \& 3, the disk blackbody is tied between the quiescent and flare states of the 2006 \XMM\ observation, while the additional hot component responsible for the QPE flares is left free to vary.}
\centering
\begin{tabular}{c c c c c c c c }
\hline
\hline
\multicolumn{7}{c}{Model 1: \texttt{tbabs$\times$ztbabs$\times$diskbb}} \\
\multicolumn{7}{c}{ $N_H = 3.0 \times 10^{20} $ cm$^{-2}$} \\
 Spectrum & $N_H(z)$ & $kT^{disk}$ & $f^{disk}_{0.5-2}$ & $\chi^2/\nu$ & \\
   & [$10^{21}$ cm$^{-2}$] & [eV] &  [erg cm s$^{-1}$] &  \\
2006 quiescent & \multirow{2}{*}{$1.6 \pm{0.3}$} & $113\pm{5.4}$ & $(7.7\pm 1.2)\times 10^{-13}$ & \multirow{2}{*}{$425/325$} & \\
2006 flare & & $178\pm{11}$  & $(1.6\pm 2.8)\times 10^{-12}$ & & \\
2021 & $0.92\pm{0.1}$ & $122\pm{29}$  & ($1.7\pm 0.2)\times 10^{-14}$ & $158/145$ &  \\
\hline
\multicolumn{7}{c}{Model 2: \texttt{tbabs$\times$ztbabs$\times$(diskbb+powerlaw)}}\\
\multicolumn{7}{c}{ $N_H = 3.0 \times 10^{20} $ cm$^{-2}$} \\
 Spectrum & $N_H(z)$ & $kT^{disk}$ & $f_{0.5-2}^{disk}$ & $\Gamma$ & $f_{0.5-2}^{pow}$ & $\chi^2/\nu$ \\
   & [$10^{21}$ cm$^{-2}$] & [eV] &  [erg cm s$^{-1}$] & & [erg cm s$^{-1}$] & \\
2006 quiescent & \multirow{2}{*}{$3.8 \pm{0.8}$} & \multirow{2}{*}{$82.9\pm{7.8}$} & \multirow{2}{*}{$(3.2\pm{0.18})\times 10^{-12}$} & --- & --- & \multirow{2}{*}{$364/323$} \\
2006 flare & & & & $4.63\pm{0.57}$ & $(2.3\pm{0.2})\times 10^{-12}$ &  \\
2021 & $3.0\pm{1.8}$ & $74 \pm{17}$ & $(5.7\pm 0.5)\times 10^{-14}$ & $1.8 \pm{1.6}$ & $(8.4 \pm 1.7)\times 10^{-15}$ & $145/145$ \\
\hline
\multicolumn{7}{c}{Model 3: \texttt{tbabs$\times$ztbabs$\times$(diskbb+bbody)}}\\
\multicolumn{7}{c}{ $N_H = 3.0 \times 10^{20} $ cm$^{-2}$} \\
 Spectrum & $N_H(z)$ & $kT^{disk}$ & $f_{0.5-2}^{disk}$ & $kT^{bb}$ & $f_{0.5-2}^{bb}$ & $\chi^2/\nu$ \\
   & [$10^{21}$ cm$^{-2}$] & [eV] &  [erg cm s$^{-1}$] & [eV] & [erg cm s$^{-1}$] \\
2006 quiescent & \multirow{2}{*}{$4.2\pm{0.9}$} & \multirow{2}{*}{$77 \pm 6.9$} & \multirow{2}{*}{$(3.4\pm{0.3})\times 10^{-12}$} & --- & --- & \multirow{2}{*}{$266/323$} \\
2006 flare & & & & $165\pm{21}$ & $(8.5\pm{1.2})\times 10^{-13}$ & \\
2021 & $4.0$ & $108\pm{16}$ & $(7.2\pm 1.1)\times 10^{-14}$ & $35\pm{10}$  & $(3.0\pm 0.5)\times 10^{-14}$ & $150/144$ \\
\hline
\multicolumn{7}{c}{Model 4: \texttt{tbabs$\times$zxipcf$\times$diskbb}}\\
\multicolumn{7}{c}{ $N_H = 3.0 \times 10^{20} $ cm$^{-2}$ ; $\log(\xi) = 2.95$} \\
 Spectrum & $N_H(z)$ & $kT^{disk}$ & $f^{disk}_{0.5-2}$ & $\chi^2/\nu$ & \\
   & [$10^{23}$ cm$^{-2}$] & [eV] &  [erg cm s$^{-1}$] &  \\
2006 quiescent & \multirow{2}{*}{$1.7.\pm{1.2}$} & $176\pm{12}$ & $(3.5 \pm 0.16) \times 10^{-13}$ & \multirow{2}{*}{$493/325$} & \\
2006 flare &  & $244\pm{18}$  & $(1 \pm 0.54) \times 10^{-12}$ & & \\
2021 & $1.8\pm{4.9}$ & $304\pm{37}$  & $(1.4 \pm 0.1) \times 10^{-14}$ & $128/165$ & \\
\hline
\hline
\end{tabular}
\label{tab:specfit}
\end{table*}

\begin{figure*}
\centering
\includegraphics[width=0.95\textwidth]{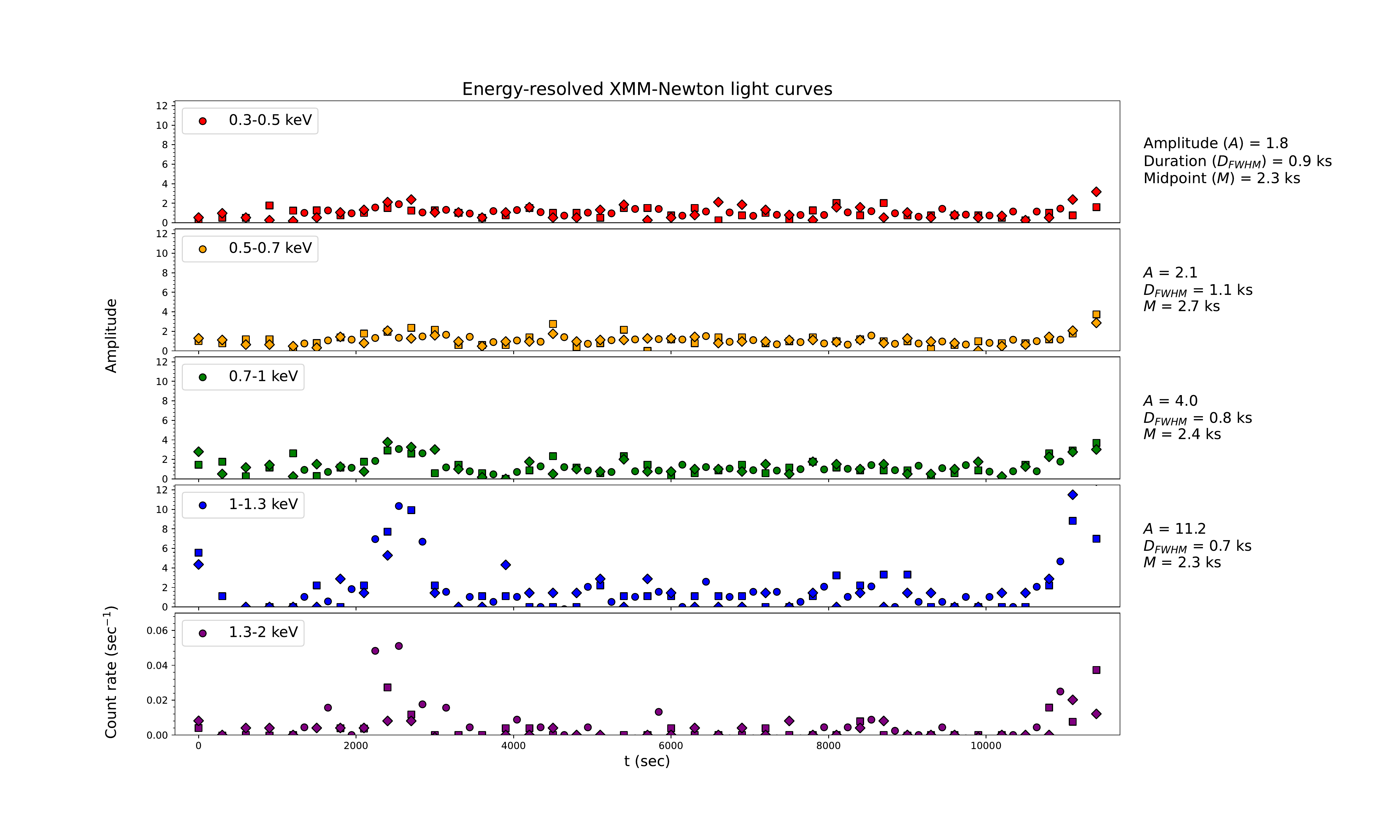}
\caption{Energy-resolved XMM light curves with time binning of 300 seconds and amplitude normalized to quiescent count rate (PN data are circles, MOS1 are squares, MOS2 are diamonds). The flares predominantly occur in the 1-1.3 keV band. We plot count rate rather than amplitude in the bottom 1.3-2 keV panel due to low S/N.}
\label{fig:eresolved}
\end{figure*}

\end{document}